\documentclass[10pt]{iopart}
\input{epsf}
\usepackage{amssymb}

\def\text#1{{\hbox{#1}}}

\def\dirac{i\partial\!\!\!\!/+eA\!\!\!\!/}

\def\lg{\lgroup}
\def\rg{\rgroup}
\def\ve{\varepsilon}

\newcommand{\beq}{\begin{eqnarray}}
\newcommand{\eeq}{\end{eqnarray}}
\newcommand{\nn}{\nonumber}

\begin{document}

\article[QHE in graphene and effective action]{}{The quantum Hall effect in graphene samples and
the relativistic Dirac effective action}

\date{\today}

\author{C.G. Beneventano}\address{Departamento de F{\'\i}sica -
Universidad Nacional de La Plata and IFLP- CONICET}
\author{Paola Giacconi}\address{Istituto Nazionale di Fisica Nucleare - Sezione di Bologna}
\author{E. M. Santangelo}\address{Departamento de F{\'\i}sica -
Universidad Nacional de La Plata and IFLP- CONICET}
\author{Roberto Soldati}\address{Dipartimento di Fisica - Universit\`a
di Bologna, Istituto Nazionale di Fisica Nucleare - Sezione di
Bologna}

\begin{abstract}
We study the Euclidean effective action per unit area and the
charge density for a Dirac field in a two--dimensional spatial
region, in the presence of a uniform magnetic field perpendicular
to the 2D--plane, at finite temperature and density. In the limit
of zero temperature we reproduce, after performing an adequate
Lorentz boost, the Hall conductivity measured for different kinds
of graphene samples, depending upon the phase choice in the
fermionic determinant.

\end{abstract}

\maketitle

Graphene is a bidimensional array of carbon atoms, packed in a
honeycomb crystal structure. Actually, each layer of a graphene
sample can be viewed as either an individual plane extracted from
graphite, or else as an array of unrolled carbon nanotubes. Stable mono-, bi- and multi-layer samples of such material have
been, recently and independently, obtained by two groups
\cite{nature1,nature2}, and a surprising behavior of the Hall
conductivity and related density of states in mono-layer samples
has been unravelled. Even more recently, an equally unexpected,
though different, behavior was reported \cite{novo2} for bi-layer
samples.

The most remarkable feature of graphene's structure, from the
theoretical point of view, is that its quanta, or
quasi--particles behave as two species (to account for the
spin of the elementary non--relativistic constituents) of massless
relativistic Dirac particles in the two non--equivalent
representations of the Clifford algebra (which correspond to the two
non--equivalent vertices in the first Brillouin zone~\cite{semenoff,guinea})
with an effective ``speed of light" about two orders of magnitude smaller than $c\,.$

To the best of our knowledge, the first approach to the quantum
Hall effect in terms of a quantum relativistic Dirac field theory, at
finite temperature and chemical potential, appeared
in reference~\cite{schakel}, where a dimensional reduction argument
was used. Later on, the relativistic Hall conductivity
was obtained in~\cite{castro,gusynhall} using Green's function methods
(for an entirely different calculation see, {\it e.g.}, \cite{ando}).
In references~\cite{schakel,castro,gusynhall}, the divergent series were regularized through definitions which, as we will explain, are equivalent to neglecting the phase of the determinant. In the zero temperature
limit, the results of refs.~\cite{schakel,castro,gusynhall} reproduce the
unexpected behavior of both the Hall conductivity and the density
of states, as measured in mono-layer graphene \cite{nature1,nature2}.

In the couple of papers \cite{BS1,BS2}, two of the authors of the
present letter developed a finite temperature field theory
calculation based upon $\zeta-$function regularization of the
Dirac determinant, and obtained the partition function and the
related Hall current and density of states. There,
the phase of the determinant was included, and its
sign fixed according to the conventional wisdom
\cite{ecz}, which lead to a Hall conductivity displaying a plateau around zero chemical potential.

It is the aim of this paper to show that, in turn, the inclusion
of the phase of determinant with opposite sign leads to a
Hall conductivity and to a zero-temperature density of states
which coincide with the ones recently reported in the case of
bi-layer graphene \cite{novo2}, while the behavior of mono-layer
graphene is reproduced when the phase is ignored, in coincidence
with the results in \cite{schakel,castro,gusynhall}.

In order to study the temperature-dependent effects for the system
at hand, we will consider the three-dimensional (3D) Euclidean
space, with metric $(+,+,+)$ and coordinates $(x,y,\tau=-it)\,,$
the $\gamma$--matrices $\gamma_i=\sigma_i,\,i=1,2,3$ and introduce
the chemical potential as an imaginary component of the gauge
potential \cite{actor}. Then, we will let the Euclidean imaginary
time coordinate vary according to $0\leq \tau\leq \beta$, where
$\beta=1/k_BT,\ k_B$ is the Boltzmann constant, and impose
antiperiodic boundary conditions on the Dirac field to reproduce
Fermi--Dirac statistics. As it is well known, another faithful
non--equivalent representation of the Clifford algebra exists in
odd dimensions, in which one of the gamma matrices changes sign
(or, equivalently, all of them do so). We will comment about the consequences of such a change of representation, wherever adequate, throughout the rest of the paper.

Once some suitable regularization has been introduced, the partition function in the grand-canonical ensemble is given by
\beq \ln{Z} \equiv \ln\, {\rm det}(D)= \ln\, {\rm det}(\dirac)\,,
\label{def} \eeq
where $-e$ is the electron charge. In order to evaluate the
partition function in the zeta regularization approach
\cite{dowker}, we must determine the spectral resolution
of the Euclidean Dirac operator, in the presence of a gauge
potential $A_{\mu}=(0 ,Bx,i{\mu}/{e})\,,$ which corresponds to the
selection of a non-symmetric gauge for the magnetic field
orthogonal to the plane -- in what follows $B>0\,.$ Here below we
just report the main results -- for a detailed calculation see,
{\it e.g.}, \cite{BS1,BS2}. The equation to be solved is
(from here on natural units are used, {\it i.e.}, $\hbar=c=1$,
unless explicitly stated)
\beq \nn [\,\sigma_1\,i\partial_x+\sigma_2\, (i\partial_y+e B x)
+\sigma_3(i\partial_{\tau}+i\mu)-\omega\,]\,\Psi=0\,.
\label{basic} \eeq
After writing
\beq \nn \Psi_{k,\,l}(x,y,\tau)
=({2\pi\beta})^{-1/2}\exp\{iky+i\tau\lambda_{\,l}\}\psi_{k,\,l}
(x)\,,\eeq where \beq \nn \psi_{k,\,l}
(x)=\left\lg\begin{array}{c}
     \varphi_{k,\,l}\,(x) \\
     \chi_{k,\,l}\,(x)  \\
\end{array}\right\rg\,,\quad
k\in{\mathbb{R}}\,, \nn\eeq
and
\beq \nn \lambda_{\,l}=(2l+1)\,({\pi}/{\beta})\,,\qquad
l\in{\mathbb{Z}}\nn \eeq 
in order to satisfy the antiperiodic boundary conditions, one
finds two types of eigenvalues and corresponding sets of eigenfunctions.

[\,I\,]\quad Asymmetric part of the spectrum

\[\omega_l=
-(2l+1){\pi}/{\beta}+i\mu, \qquad l=-\infty,...,\infty\]

The corresponding eigenfunctions are quite particular, in that they are eigenfunctions of $\sigma_3$ with eigenvalue $+1\,,$ the ones in the orthogonal subspace being indeed eliminated by the square-integrability condition (for some related references see, for instance, \cite{klimenko}). As a consequence, the
corresponding eigenvalues $\omega_l$ are not the square roots of
the eigenvalues of $D^{\dagger}D$. They will eventually lead to a
spectral asymmetry and, thereby, to a phase of the determinant,
which will be studied in detail below.

[\,II\,]\quad Symmetric part of the spectrum

\[\omega_{l,\,n}^{\pm}=\pm \sqrt{{\tilde
\lambda_l}^2+2n e B}, \qquad n=1,...,\infty,
\, l=-\infty,...,\infty\]

For both kinds of eigenfunctions, the degeneracy per unit area is given
by the well known Landau factor $\Delta_L={eB}/{2\pi}\,.$

It is worthwhile to remark that, had we chosen the other
non--equivalent representation of the $\gamma$--matrices, the
eigenvalues of type [\,I\,] would have changed their sign, and the
corresponding eigenfunctions would have been eigenfunctions of $\sigma_3$ with eigenvalue $-1$. However, as will be discussed below, this fact will not
lead to any modification in our physical predictions as long as $\mu$ is
real. Thus, considering the contributions of both nonequivalent
representations will amount to an overall factor of two.

When parity is defined as, {\it e.g.}, in ref.~\cite{dunne},
it is easy to check that, for a general
Dirac operator, the effect on the spectrum is
$\omega^{P}=-\omega$. This symmetry is obviously respected by the
symmetric part [\,II\,] of our spectrum, while it actually
produces a change in the sign of the asymmetric portion [\,I\,].
When acting on the latter, it is equivalent to $\mu\rightarrow
-\mu$ and, thus, to charge conjugation ($\psi(x)\rightarrow
\gamma_2\,{\psi}^{*}(x);\, A_\nu(x)\rightarrow -A^{\,*}_\nu(x);\,
\omega\rightarrow \omega^{C}={\omega}^{*}$). So, parity is broken,
already at the {\it classical} level if only one representation of
the gamma matrices is considered, due to the square--integrability
condition.

If complex values of $\mu$ are allowed, the whole spectrum has an
interesting symmetry: it turns out to be invariant under
$\mu\beta\rightarrow \mu\beta+{2\pi ik}\,,\ k\in\mathbb{Z}\,,$
which is nothing but the symmetry under ``large" gauge
transformations. The conflict between this last symmetry and
parity invariance in different regularization schemes is well
known \cite{poly}, and created some controversy in the past
\cite{dunne,deser,schapo}.

In this letter, we will concentrate on the case of a real
chemical potential $\mu$. A discussion of both symmetries
in the case of a complex chemical potential will be reported
elsewhere~\cite{complex}.

Starting from the above described spectrum, we
shall evaluate, according to (\ref{def}) the Euclidean effective
action per unit area (in the
statistical mechanics terminology, the latter coincides with the
grand-potential per unit area in units of $k_BT$). From this
effective action, the mean fermionic number per unit area,
$N_{\,\rm F}$, and, thus, the charge density can be retrieved as
follows
\beq S_{eff}=\log{Z}\equiv\ln\, {\rm det}(D)\,,\quad N_{\,\rm
F}={\beta}^{-1}\frac{\partial S_{eff}}{\partial\mu}\,.
\label{numero}\eeq

Here, the symbol $\equiv$ stands for the definition through an adequate regularization. For the reasons we have just explained
({\it viz.}, all the eigenvalues are paired), it turns out that
the contribution to
the effective action coming from the symmetric part
[\,II\,] of the spectrum does not suffer from regularization
ambiguities. For instance, after a proper definition in terms of
the $\zeta$--regularization \cite{BS1}, it is given by
\beq \fl S_{eff}^{II}
= \Delta_L\left[\beta\,\zeta_R(-{1}/{2})\,(2e B)^{1/2}
+  \sum_{n=1}^{\infty} \log\left\{\left(1+z\,e^{-\beta
{\ve}_n}\right) \left(1+z^{-1}e^{-\beta
{\ve}_n}\right)\right\}\right] \label{Fmenor} \eeq
with $z:=\exp\{\beta\mu\}\,,\ \ve_n\equiv\sqrt{2neB}\,$.

The contribution to the effective action arising from the
asymmetric part [\,I\,] of the spectrum is given by the formal
expression \beq \fl S_{eff}^{I}=
\Delta_L\sum_{l=0}^{\infty}\log\left\{(-1)[\,(2l+1){\pi}/{\beta}-i\mu\,] \times [\,(2l+1){\pi}/{\beta}+i\mu\,]\right\}\,.
\label{formal} \eeq
Choosing a symmetric regularization, as done in
\cite{schakel,castro,gusynhall}, is equivalent to ignoring the infinite
term $\sum_{l=0}^{\infty}\log(-1)\,,$ which reduces the previous
expression to
\beq S_{eff}^{I}=\Delta_L
\sum_{l=0}^{\infty}\log\left\{[\,(2l+1){\pi}/{\beta}\,]^2 +
{\mu}^2\right\}\,,\nn \eeq {\it i.e.}, one evaluates the logarithm
of the ``absolute value" of the Euclidean Dirac operator which,
once regularized, leads to
\beq S_{eff}^{I}=\Delta_L
\log{\left[\,2\cosh({\mu\beta}/{2})\,\right]}\,. \label{nophase}
\eeq
However, a first--principle $\zeta$--function regularization of
the determinant unavoidably drives to a careful definition of the
phase of the determinant, which is equivalent to the selection of a cut in
the complex plane of the eigenvalues \cite{quique2}, when the asymmetric
part of the spectrum is treated. Going back to
the formal relation (\ref{formal}), we define it in proper
mathematical sense in terms of the $\zeta$--function
regularization: namely,
\beq \fl \left. S_{eff}^{I}:= -\Delta_L
\frac{d}{ds}\right\rfloor_{{s=0}}\,
\left\{\sum_{l=0}^{\infty}\left[\,(2l+1)\,
\frac{\pi}{\beta}+i\mu\,\right]^{\,-s}+ \sum_{l=0}^{\infty}\left[\,e^{\pm\pi i}(2l+1)\,
\frac{\pi}{\beta}+i\mu\,\right]^{-s}\right\} \,,\nn \eeq
the phase of the determinant being fixed by the cut in the complex
plane of the eigenvalues \cite{quique}. More explicitly we can
write
\beq \fl S_{eff}^{I} :=\left.
-\Delta_L\,\frac{d}{ds}\right\rfloor_{{s=0}}\,
\left\{\sum_{l=0}^{\infty}\left[\,(2l+1)\,
\frac{\pi}{\beta}+i\mu\,\right]^{-s}+\sum_{l=0}^{\infty}e^{-is\theta}\left[\,(2l+1)\,\frac{\pi}{\beta}
+i\mu\,e^{-i\theta}\,\right]^{-s}\right\}\ ,\nn \label{phase} \eeq
where $-\pi\le\theta\le\pi\,.$ The prescription usually adopted
\cite{ecz} amounts to choosing the cut in such a way that the
expression in the last square bracket never vanishes, as one goes
continuously from the eigenvalues with a positive real part to
the eigenvalues with a negative real part, {\it i.e.},
$(2l+1){\pi}/{\beta}+\mu\sin{\theta}$ and $\mu\cos{\theta}$ do not
simultaneously vanish (which could happen if
$\mu=(2l+1){\pi}/{\beta}$ and $\theta=-{\pi}/{2}$ or
$\mu=-(2l+1){\pi}/{\beta}$ and $\theta={\pi}/{2}$). This requires
that the cut is chosen below (above) the real axis when $\mu$ is
positive (negative), which gives for the final value
$\theta=\pi{\rm sign}\mu$. With this choice, the contribution of
the asymmetric part of the spectrum -- see ref.~\cite{BS1} for
more details -- to the effective action is given by
\beq S_{eff}^{I}&=&
\Delta_L\,\left[-\frac12\,\beta\,|\mu|+\log\left(2\cosh\frac{\mu\beta}{2}\right)\right]\,.
\label{usualphase} \eeq
The opposite, and less popular, definition of the phase would lead
to
\beq S_{eff}^{I}&=&
\Delta_L\,\left[\frac12\,\beta\,|\mu|+\log\left(2\cosh\frac{\mu\beta}{2}\right)\right]\,.
\label{unusualphase} \eeq
\medskip

At this point, it is important to stress (always in the case of
a real chemical potential) that exactly the same results are obtained,
provided the same criteria are applied, if the other non--equivalent
representation for the $\gamma$--matrices is chosen. Thus, the inclusion of
both contributions, amounts to an overall factor of two, if the
phase is consistently chosen in both representations. In this
sense, the exclusion of the phase is equivalent to the adoption of
opposite criteria for the phase selection in both representations.

Putting together the contributions from the symmetric part
(\ref{Fmenor}) and the asymmetric part of the spectrum
(\ref{nophase}), (\ref{usualphase}) or (\ref{unusualphase}),
depending on the phase definition adopted, we come to the following expression for the $\zeta$--function definition of the Euclidean effective action,
\beq
 S_{eff}\ &=&\Delta_L\,\left\{\log\left(2\cosh\frac{\mu\beta}{2}\right)\right.
+
\frac12\,\kappa\,\beta\,|\mu|+\beta\,\zeta_R(-{1}/{2})\sqrt{2e B}
\nonumber\\
& +&\left.\sum_{n=1}^{\infty}
\log{\left[\left(1+z\,e^{-\beta\ve_n}\right)
\left(1+z^{-1}e^{-\beta\ve_n}\right)\right]}\right\},\nn \\
&& \kappa= 0,\pm 1\,,\quad z=\exp\{\beta\,\mu\}\,,\quad
\ve_n=\sqrt{2neB}\ . \nn\label{accion} \eeq
Here, $\kappa=0$ corresponds to a vanishing phase, $\kappa=-1$ to
the usual phase choice and  $\kappa=+1$ to the opposite and
unusual phase choice. Note that $\ve_n$ is the absolute value of
the $n-$th non--vanishing Landau level. In all cases, the
Euclidean effective action is an even function of $\mu\,$. Thus,
it is invariant under charge conjugation and parity.

Also in all the cases, the mean fermionic number per unit
area, turns out to change sign under $\mu\rightarrow -\mu\,$. In
fact, from its very definition in (\ref{numero}), we get
\beq \fl
N(\kappa\,;\beta,\mu)\ &=&{\Delta_L}\left\{
\frac12\tanh\frac{\mu\beta}{2} + \frac12\kappa\,{\rm sgn}(\mu)\right.\nn \\ \fl
&+& \!\!\sum_{n=1}^{\infty}\ \left[\,1+e^{\beta\sqrt{2ne
B}-\mu\beta}\,\right]^{-1} - \left.
\!\!\sum_{n=1}^{\infty}\ \left[\,1+e^{\beta\sqrt{2ne
B}+\mu\beta}\,\right]^{-1}\right\}\,.
\nonumber
\label{Nmenor}
\eeq
Note that the first two terms are the ones coming from the
asymmetric part of the spectrum. So, no matter how one defines the
phase of the determinant, at any finite temperature there is a
kind of parity breaking charge, which is the sum of a $\mu$-analytic
contribution (the first term) and a non-analytic one (the phase
of the determinant).
In the zero temperature limit, for $n<\mu^2/2eB<n+1\,,$ we finally
obtain
\beq \lim_{\beta\to\infty}\,N(\kappa\,;\beta,\mu)\ =\
\left(n+\frac{1+\kappa}{2}\right)\,{\rm sgn}(\mu)\,\Delta_L\,.
\label{fermionico}
\eeq

In this limit, the contribution of the asymmetric part of the
spectrum is the one corresponding to $n=0$, and is non--analytic
in all cases. It only vanishes, so that parity and charge
conjugation symmetries are indeed fulfilled, for the most commonly
accepted selection of the phase $\kappa=-1\,.$

Moreover, it is easy to check that Nernst's theorem holds true for any $\kappa\,.$
Actually, it turns out that the entropy
can be obtained
from the well known Boltzmann--von Neumann formula
\beq
S(\beta,\mu;B,\kappa)=k_B\left(1-\beta\,
{\partial\over\partial\beta}\right)S_{eff}\,;
\nn \eeq
whence, one can verify by direct inspection that, indeed,
\beq
\lim_{\beta\to\infty}S(\beta,\mu;B,\kappa)\ =\ 0\,,\qquad
\forall\kappa=0,\pm 1\,,\nn \eeq
in agreement with Nernst's theorem.

From (\ref{fermionico}), the mean value of the charge density
in the zero temperature limit can be immediately obtained
for one representation and one fermion species.
Turning back to physical units, and recalling that the particle charge
is $-e\,,$ we find
\beq
J_{\,\tau}(\kappa\,;\beta,\mu)\ &=&\
-\,eN_{\,\rm F}(\kappa\,;\beta,\mu)\nn\\
&{\buildrel\beta\to\infty\over \longrightarrow}&\
-\,{\rm sgn}(\mu)\,
\left(n+\frac{1+\kappa}{2}\right)\,{e^2 B\over hc}\,,\nn \\
&{\rm for}& n<({\mu^2}/{2eB\hbar c^2})<n+1\,,\
n\in{\mathbb N}\,,\nn
\eeq
the spatial components of the current density being equal
to zero in the absence of electric fields. Now, the zero
temperature limit of the same vector in the presence of
crossed homogeneous electric $E^{\,\prime}$ and magnetic
$B^{\,\prime}$ fields can retrieved, for $E^{\,\prime}<
B^{\,\prime}$, by performing a Lorentz boost with absolute value
of the velocity $v=c{E^{\,\prime}}/{B^{\,\prime}}\,.$ Suppose, for
definiteness, that the homogeneous electric field points toward
the positive $Oy$--axis. Then, the speed of the Lorentz boost
must point toward the negative $Oy$--axis and the transformation
law gives as a result
\beq
&& J_{\,\tau}^{\,\prime}(\kappa\,;\mu)\ =\ -\,{e^2 B^{\,\prime}\over hc}\,
\left(n+\frac{1+\kappa}{2}\right)\,{\rm sgn}(\mu)\,,\nn \\
&& J_{\,x}^{\,\prime}(\kappa\,;\mu)\ =\ -\,{e^2 E^{\,\prime}\over h}\,
\left(n+\frac{1+\kappa}{2}\right)\,{\rm sgn}(\mu)\,,\nn \\
&& J_{\,y}^{\,\prime}(\kappa\,;\mu)\ =\ 0\,,
\nn \eeq
for $n\,<\,({\mu^2}/{2eB\hbar c^2})\,<\,n+1\,,\ n\in{\mathbb N}\,.$
As a consequence, the contribution to the quantized Hall conductivity
at zero temperature becomes, for each representation and each fermion species,
\beq
\sigma_{xy}\ =\ -\,{e^2\over h}\left(n+\frac{1+\kappa}{2}\right)\,{\rm sign}(\mu)\,,
\qquad  n\in{\mathbb N}\,.
\nn \eeq
As explained throughout the paper, this result must be multiplied
by an overall factor of four, in order to make contact with the
relativistic effective theory associated to graphene \cite{semenoff,guinea}. It is interesting to remark that the three values of the phase correspond to three different vacuum polarization (Casimir-type) effects due to the interaction with the magnetic field at zero temperature. More precisely, the quantum (in the field-theory sense) filling factor is given, in each of the three cases, by

\beq
\nu_Q=-\frac{J_{\,\tau}^{\,\prime}h\,c}{e^2 B^{\prime}}=\left(n+\frac{1+\kappa}{2}\right)
\eeq

\begin{figure}

\epsffile{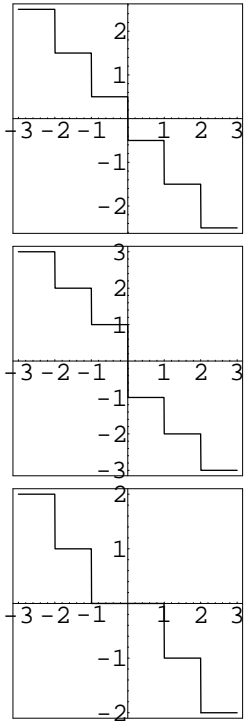}
\caption{Hall conductivity for different selections of the phase of the determinant.
Top to bottom: $\kappa=0\,,$ $\kappa=1\,,$ $\kappa=-1\,.$
In all cases, the horizontal axis represents $\nu_F={\rm sgn}(\mu)\,\mu^2/{2eB\hbar c^2}\,$
and the vertical one, ${\sigma_{xy}\,h}/{4e^2}\,.$}

\label{figure}
\end{figure}

The (rescaled) Hall conductivity is presented in figure \ref{figure}, for the three values of $\kappa\,$, as a function of $\nu_C={\rm sgn}(\mu)\,\mu^2/{2eB\hbar c^2}$, which is nothing but the classical density of carriers in the relativistic theory, divided by the total degeneracy of each Landau level.

It is well known that any regularization procedure is acceptable,
unless either it  manifestly violates some of the symmetries of the system,
or it is ruled out by the experimental data. So, the measured Hall conductivities of graphene as reported, for instance, in reference \cite{novo2}, should shed light on the relevance (or lack thereof) of the different phases of the determinant.

On these grounds, the first and clearer conclusion of this letter is that the behavior of monolayer graphene, as presented not only in \cite{novo2} but also in \cite{nature1,nature2}, corresponds to $\kappa=0$, i.e., to not including the phase of the determinant, as done in \cite{schakel,castro,gusynhall}. In fact, in this case the (rescaled) Hall conductivity shows a jump of height $1$ for $\nu_C=0$, and further jumps of the same magnitud for $\nu_C=\pm1,\pm2,...$.

Let us now compare our predictions with the contents of reference \cite{novo2}, which is devoted to bilayer graphene. In this case, the (rescaled) Hall conductivity presents a jump of height $2$ for $\nu_C=0$, and further jumps of height $1$. The main point here concerns the positions of these subsequent jumps. As a matter of fact, according to figure 1.b in the same reference, these subsequent jumps appear for $\nu_C=\pm1,\pm2,...$, which is exactly the behavior predicted, in our calculation, for $\kappa=+1$ (the less popular selection of phase in the Dirac determinant). However, the same reference interprets the Hall behavior of bilayer graphene through a theoretical prediction first made in \cite{falko}, where the theory is ``almost" nonrelativistic, with a Landau spectrum given by $E_n=\pm\frac{eB\hbar}{m}\sqrt{n(n-1)}$. This last model does, indeed, predict a double jump for $\nu_C=0$, due to the existence of two zero modes in each representation. But, for the very same reason, the next jump should appear, in this theoretical scenario, at $\nu_C=\pm\sqrt{2}\sim\pm\frac32$ (for a related discussion see, for instance, \cite{kope}). As stressed by the authors, figure 1.b in \cite{novo2} is only schematic. However, the experimental results corresponding to $B=12T$ in figures 2.b and 2.c of the same reference also tend to confirm the prediction of our relativistic quantum field calculation, with $\kappa=1$ (unusual phase), where the next jumps occurs at $\nu_C=\pm1$. The text in the same reference also seems to confirm our prediction, since the distance between jumps is said to be, always for $B=12T$, $\Delta n\sim 1.2\times 10^{12}cm^{-2}$, which corresponds to $\Delta \nu_C=\Delta n \frac{h}{4eB}\sim 1$, the same for all jumps. However, the experimental results corresponding to $B=20T$ seem to agree with the width of the first plateau being approximately 1.3 times the width of the subsequent ones. Thus, further experimental study of bilayer graphene is crucial in distinguishing between both theoretical scenarios.

At this point, one can naturally wonder whether there is place at all for the usual selection of phase ($\kappa=-1$) in the description of graphene samples.
It is quite interesting to gather that the three non--equivalent
phase selections correspond to the three non--equivalent unitary representations of the cyclic group $C_3$, which is precisely the relevant symmetry
group for graphene. Even though the study of the Hall conductivity in graphene samples with three layers \cite{berger} is certainly far from being conclusive, they seem to indicate that a quantum Hall effect does occur in such devices, with a plateau at $\nu_C=0$. Does the usual phase selection correspond to the behavior of three-layered graphene?

In any case, from a theoretical point of view, further experiments on graphene samples can give an answer to a long-standing question in the field of the zeta-function regularization, i.e.: Which phase must be selected in the definition of Dirac determinants, in order to evaluate effective actions?

\begin{ack}

{\em Acknowledgements} : C.G.B. and E.M.S. would like to
thank Alejandro Cabrera, Horacio Falomir and Yakov Kopelevich for useful discussions and suggestions.
The work of C.G.B. and E.M.S. was partially supported by UNLP (Proyecto 11/X381) and CONICET (PIP 6160).

\end{ack}

\end{document}